\documentstyle[preprint,revtex]{aps}
\begin{document}
\draft
\preprint{UAHEP-9213}
\preprint{July 1992}
\begin{title}
Dilatonic Black Holes, Naked Singularities and Strings
\end{title}
\author{P. H. Cox\cite{cox}, B. Harms and Y. Leblanc}
\begin{instit}
Department of Physics and Astronomy,
The University of Alabama\\
Box 870324,
Tuscaloosa, AL 35487-0324
\end{instit}
\begin{abstract}
We extend a previous calculation which treated Schwarzschild
black hole horizons as quantum mechanical objects
to the case of a charged,
dilaton black hole.  We show that for a unique value of the
dilaton parameter $a$, which is determined by the condition
of unitarity of the S matrix, black holes transform at the
extremal limit into strings.
\end{abstract}
\narrowtext

Several authors \cite{thooft,verl,holz,witten} have discussed the
relationship between strings and the singularities encountered in
general relativity.  In Refs.  \cite{thooft,verl} string-like
actions were derived from two different, although presumably
equivalent, analyses of the scattering of massless, point-like
particles from a Schwartzschild black hole.  Holzhey and Wilczek
\cite{holz} have suggested that extremal charged, dilatonic
black holes may evolve from membranes to strings when the dilaton
parameter increases to one.  Witten \cite{witten} has pointed out
that the ``cosmic censorship'' hypothesis may be wrong and that
string theory may provide useful insights into the study of naked
singularities.

In this letter we analyze the scattering of a massless,
pointlike particle from a charged, dilatonic black hole.  We
obtain the dependence of the phase shift and the string tension
on the dilaton parameter $a$.  We show that for the extremal,
dilatonic black hole there is a value of $a$ for which a black
hole transforms into a string.

In Ref. \cite{dray} the phase shift of a massless particle
scattering from a Schwarzschild black hole was calculated.
't Hooft \cite{thooft} showed that under rather general
assumptions about quantum gravitodynamics the S matrix which
arises from this phase shift can be written in terms of a
``string-like'' action with imaginary string tension equivalent.
We first extend this calculation by
deriving the Green function which determines the phase
shift for a massless particle scattering from a
charged, dilatonic black hole.  As in Ref. \cite{dray},
the metric of space-time
surrounding a black hole with a massless particle falling in
from spherical angle $\Omega' = (\theta' , \phi')$ can under
certain conditions be represented by gluing together
two solutions of Einstein's field equations.  The solutions are
cut at the null surface defined by setting one of the
Kruskal coordinates to zero ($u = 0$) and glued after a
nonconstant shift of the other Kruskal coordinate,
\begin{eqnarray}
\delta v(\Omega) = p_{in} f(\Omega,\Omega ') \,
\end{eqnarray}
where
$p_{in}$ is the ingoing particle's momentum with respect to the
Kruskal coordinates
and the Green function $f$ satisfies the
equation , at $u=0$,
\begin{eqnarray}
{A\over{g}}\Delta f - {g_{,uv}\over{g}} f = 32 \pi A^2
\delta^2 (\Omega,\Omega ') \big|_{u=0} \; ,
\end{eqnarray}
where $\Delta$ is the angular Laplacian.
The functions $A, g$ and $g_{,uv}$ are identified by writing the
metric in the form
\begin{eqnarray}
ds^2 = 2\;A(u,v)\; dudv + g(u,v)\; (d\theta^2 + \sin^2\theta
d\phi^2)\; .
\end{eqnarray}
The necessary conditions for the matching are
\begin{eqnarray}
A_{,v}\big|_{u=0} = g_{,v}\big|_{u=0} = 0\; .
\end{eqnarray}

To find $A(u,v)$ and $g(u,v)$ we start from the metric for a
charged, dilaton black hole written in terms of the usual
space-time coordinates
\begin{eqnarray}
ds^2 = \lambda^2\; dt^2 - \lambda^{-2}\; dr^2 - R^2\; (d\theta^2
+ \sin^2 \theta\; d\phi^2),
\end{eqnarray}
where ($a$ is the dilaton parameter)
\begin{eqnarray}
\lambda^2 = (1-{r_+ \over{r}})\; (1-{r_- \over
{r}})^{(1-a^2)/(1+a^2)} \; ,
\end{eqnarray}
and
\begin{eqnarray}
R^2 = r^2\; (1- {r_-\over{r}})^{2a^2/(1+a^2)} \;
\end{eqnarray}
The parameters $r_+$ and $r_-$ are related to the mass, $M$, and
the charge, $Q$, by
\begin{eqnarray}
2\;M &=& r_+ + {(1-a^2)\over{(1+a^2)}}\; r_-\nonumber \\
Q^2 &=& {r_+\; r_- \over{ 1+a^2}} \; .
\end{eqnarray}
The Kruskal coordinates are obtained as usual (see, e.g.,
Ref. \cite{adler}).  $u$ and $v$ are related to the
time, $t$,
and the ``tortoise'' coordinate, $\xi$, by
\begin{eqnarray}
u &=& e^{\alpha \xi}\; e^{\alpha t} \nonumber \\
v &=& -e^{\alpha \xi}\; e^{-\alpha t}\; ,
\end{eqnarray}
where $\alpha$ is a constant whose value will be determined by
the conditions stated in Eq.(4) and $\xi$ is given by
\begin{eqnarray}
\xi = \int {dr \over {\lambda^2}}\; .
\end{eqnarray}
Using the transformations in Eq.(9), we find
\begin{eqnarray}
A(u,v) &=& -{\lambda^2 e^{_2\alpha \xi} \over{2 \alpha^2}}
\nonumber \\
g(u,v) &=& R^2 \; .
\end{eqnarray}

In order for $A_{,v}$ and $g_{,v}$ to vanish at $u = 0\ (r =
r_+)$ we must have
\begin{eqnarray}
{d\lambda^2 \over{dr}}\bigg|_{r=r_+} -2\; \alpha  =0,
\end{eqnarray}
which gives for $\alpha$
\begin{eqnarray}
\alpha = {1 \over{2\; r_+}} \left({r_+ - r_- \over
{r_+}}\right)^b \; ,
\end{eqnarray}
where $b = (1-a^2)/(1+a^2)$.

We are now ready to determine the Green function $f$.  We
multiply through Eq.(2) by $g$ and use
\begin{eqnarray}
\Gamma =
{g_{,uv}\over{A}}
\bigg|_{u=0} \; = \;  {r_+ -r_- \over{r_+}} + {a^2\;
r_- \over{r_+(1+a^2)}}
\end{eqnarray}
If we take the north pole as the direction of incidence,
Eq.(2) becomes,
\begin{eqnarray}
\Delta f -\Gamma \; f = -2\; \pi\; \kappa\;\delta (\theta)\; ,
\end{eqnarray}
where we have defined $\kappa$ to be
\begin{eqnarray}
\kappa = - 16\; Ag\big|_{u=0} \; .
\end{eqnarray}
Evaluating $\kappa$ we find
\begin{eqnarray}
\kappa = 32\; e^{-1}\; r_{+}^2\;(1-{r_-
\over{r_+}})^{2a^2/(1+a^2)} \; ,
\end{eqnarray}
where the constant of integration has been chosen such that the
``tortoise'' coordinate reduces in the limit $ a \to 0$ to the
form obtained for the Schwarzschild solution.

The solution of Eq.(15) proceeds the same as in
Ref.\cite{dray}.  The Green function is expanded in spherical
harmonics and upon determining the expansion coefficients it can
be expressed as a sum over Legendre polynomials
$P_l(\cos\;\theta)$,
\begin{eqnarray}
f = \kappa \sum_l\; {l+{1\over {2}} \over{l(l+1)+\Gamma}}\;
P_l(\cos\theta)\; ,
\end{eqnarray}
where $\Gamma$ is the constant defined in Eq.(14).

The calculation of the S matrix \cite{thooft,verl} gives
\begin{eqnarray}
<u(\Omega)\mid v(\Omega)> = N exp\left( i \int
f^{-1} (\Omega, \Omega ')\; v(\Omega)\; u(\Omega)\right)\; ,
\end{eqnarray}
where $f^{-1}$ is $(\Gamma - \Delta)/2\pi \kappa$ and $N$ is a
normalization constant.  In the momentum representation this
becomes
\FL
\begin{eqnarray}
<p_{out}(\Omega)\mid p_{in}(\Omega)>& = &N'\;\int Du(\Omega)\;
\int Dv(\Omega)\nonumber \\
&\times& exp\left[\int
d^2\Omega\left({i\over{2\pi\kappa}}(\Gamma\;u\;v +
\partial_{\Omega}v \partial_{\Omega}u) +
iu\;p_{out}-iv\;p_{in}\right)\right]\; .
\end{eqnarray}
neglecting the term due to the curvature ($\Gamma$ term),
switching to
the `membrane coordinates' $x^0, x_3$ defined by (in the metric
where $x^2 = {\bf x}^2 - x_0^2$)
\begin{eqnarray}
x^0 = (v + u)/2 \; ;\ \ x_3 = (v-u)/2 \; ,
\end{eqnarray}
and writing the result as a covariant expression gives
\begin{eqnarray}
<p_{out}(\Omega)\mid p_{in}(\Omega)>& = &
C \int Dx^{\mu}(\sigma)\;
Dg^{ab}(\sigma)\nonumber \\
&&\times exp\left[\int
d^2\sigma\left({-T\over{2}}\sqrt{g}\;g^{ab}\;\partial_a x^{\mu}
\;\partial_b x^{\mu} +
ix^{\mu}\;p^{\mu}(\sigma)\right)\right]\;,
\end{eqnarray}
where $T$, the string tension equivalent, is
\begin{eqnarray}
T = {i \over{\pi\; \kappa}}\; ,
\end{eqnarray}
and the integration $d^2\sigma,
(\sigma_1 = x_1,\
\sigma_2 =x_2)$ is over the variables orthogonal
to the membrane coordinates generated by $\theta$ and $\phi$.
The argument of the exponential in
Eq.(22) is the Polyakov action except for the factor
of $i$.  To remove the factor of $i$ and thus restore unitarity
we can consider the case where $r_- > r_+$ and set the dilaton
parameter $a = 1/\sqrt{3}$.  In this region $\kappa$ is
pure imaginary, and
the action is truly of the Polyakov form.

Of course, the case $r_- > r_+$ represents a naked
singularity classically,
so the picture indicated by this calculation is
that charged, dilaton black holes undergo a transition as their
parameters evolve to
the extreme case (here $r_+ = r_-$) where they
would become naked singularities, instead become strings, for
this specific value of the dilaton
parameter $a$.  Presumably the event horizon then becomes
the world sheet of the string.

In this paper we have shown that 't Hooft's idea that the
horizon of a black hole may be regarded as a quantum mechanical
object can be extended to the case of charged, dilaton black
holes.  Using the constraint of unitarity, we have calculated a
unique value, $a = 1/\sqrt{3}$,
for the dilaton parameter.  We have found
that for this value black holes undergo a transition to true
strings at the extremal limit for a particular value of the
dilaton parameter.  We are exploring the possibility
that similar transitions may occur between black p-branes and
strings \cite{horo}.  In this case the unitarity constraint is
expected to give a dilaton parameter dependent upon the
space-time dimensionality.
Our considerations have been for quantum
mechanical objects which undergo a transition into strings
rather than producing a naked singularity.
It is tempting to speculate that this represents a new way of
implementing the
``cosmic censorship'' hypothesis in such processes as
aspherical gravitational collapse, which has been suggested as
a possible explanation of the observed extragalactic
gamma ray bursts. This paper shows that instead of a naked
singularity fundamental strings may be produced whose massless
modes are
detected on earth as gamma rays.  If appropriate detectors can
be
designed and constructed, the massive string modes might also be
detected for such an event.

\acknowledgements

This research was supported in part by the U.S. Department of
Energy under Grant No. \ DE-FG05-84ER40141 and in part by the
Texas National Research Laboratory Commission under Grant No.
RCFY92-117.


\begin{references}
\bibitem[\dag]{cox}
Permanent address: Department of Physics, Texas A\&I
University, Kingsville, TX 78363.
\bibitem{thooft} G. 't Hooft, Nuc.\ Phys.\ {\bf B335}, 138(1990).
\bibitem{verl} H. Verlinde and E. Verlinde, PUPT-1279\ (1991).
\bibitem{holz} C. Holzhey and F. Wilczek, IASSNS-HEP-91/71\
(1991).
\bibitem{witten} E. Witten, IASSNS-HEP-92/24 \ (1992).
\bibitem{dray} T. Dray and G. 't Hooft, Nuc.\ Phys.\ {\bf B253},
173(1985).
\bibitem{adler} R. Adler, M. Bazin, and M. Schiffer, {\it
Introduction to General Relativity}, McGraw-Hill, New York
(1965).
\bibitem{horo} G.T. Horowitz and A. Strominger, Nuc.\ Phys.\
{\bf B360}, 197(1991).
\end{references}
\end{document}